\documentclass[twocolumn,showpacs,aps,prl,superscriptaddress]{revtex4}

\usepackage{graphicx}
\usepackage{dcolumn}
\usepackage{amsmath}
\usepackage{epsfig}

\input{pubboard/babarsym}

\newcommand{\BABARPubYear}    {10}
\newcommand{\BABARPubNumber}  {009}

\newcommand{\SLACPubNumber} {14143}
\newcommand{\LANLNumber} {1005.5190}

\def\figurebox#1#2#3{
    \def\arg{#3}
    \ifx\arg\empty
    {\hfill\vbox{\hsize#2\hrule\hbox to #2{\vrule\hfill\vbox to #1{\hsize#2\vfill}\vrule}\hrule}\hfill}
    \else
    {\hfill\epsfbox{#3}\hfill}
    \fi}

\long\def\inst#1{\par\nobreak\kern 4pt\nobreak
    {\it #1}\par\vskip 10pt plus 3pt minus 3pt}

\newcommand{\x}{$X~\mathrm{meson}$}
\newcommand{\y}{$Y~\mathrm{meson}$}

\newcommand{\bpto}{$B^+\rightarrow J/\psi\pi^+\pi^-\pi^0K^+$}

\newcommand{\jo}{$J/\psi\omega$}
\newcommand{\jthreepi}{$J/\psi\pi^+\pi^-\pi^0$}
\newcommand{\mjo}{$m_{J/\psi\omega}$}
\newcommand{\mppp}{$m_{3\pi}$}
\newcommand{\xto}{$X\to J/\psi\omega$}
\newcommand{\yto}{$Y\to J/\psi\omega$}
\newcommand{\threepi}{$\pi^+\pi^-\pi^0$}

\newcommand{\carg}{$C_{\mathrm{ARG}}$}

\begin{document}

\preprint{\babar-PUB-\BABARPubYear/\BABARPubNumber} 
\preprint{SLAC-PUB-\SLACPubNumber}

\begin{flushleft}
\babar-PUB-\BABARPubYear/\BABARPubNumber\\
SLAC-PUB-\SLACPubNumber\\

hep-ex/\LANLNumber\\  [10mm]

\end{flushleft}

\title{ {\large {\bf \boldmath Evidence for the decay ${X(3872)
\rightarrow J/\psi\omega}$ } } }

%% author list as of 04-Apr-2010 (440 authors)
%
\author{P.~del~Amo~Sanchez}
\author{J.~P.~Lees}
\author{V.~Poireau}
\author{E.~Prencipe}
\author{V.~Tisserand}
\affiliation{Laboratoire d'Annecy-le-Vieux de Physique des Particules (LAPP), Universit\'e de Savoie, CNRS/IN2P3,  F-74941 Annecy-Le-Vieux, France}
\author{J.~Garra~Tico}
\author{E.~Grauges}
\affiliation{Universitat de Barcelona, Facultat de Fisica, Departament ECM, E-08028 Barcelona, Spain }
\author{M.~Martinelli$^{ab}$}
\author{A.~Palano$^{ab}$ }
\author{M.~Pappagallo$^{ab}$ }
\affiliation{INFN Sezione di Bari$^{a}$; Dipartimento di Fisica, Universit\`a di Bari$^{b}$, I-70126 Bari, Italy }
% ...
\collaboration{The \babar\ Collaboration}
\noaffiliation

\date{\today}
\begin{abstract}
We present a study of the decays $B^{0,+}\to J/\psi\pi^+\pi^-\pi^0
K^{0,+}$, using 467$\times 10^6$ $B\bar{B}$ pairs recorded with the
\babar\ detector. We present evidence for the decay mode $X(3872)\to
J/\psi\omega$, with product branching fractions ${\cal{B}}(B^+\to
X(3872)K^+)\times {\cal{B}}(X(3872)\to J/\psi\omega$)
$=[0.6\pm0.2\stat \pm 0.1\syst ]\times 10^{-5}$, and ${\cal{B}}(B^0\to
X(3872)K^0)\times {\cal{B}}(X(3872)\to J/\psi\omega$)
$=[0.6\pm0.3\stat \pm 0.1\syst ]\times 10^{-5}$. A detailed study of
the $\pi^+\pi^-\pi^0$ mass distribution from $X(3872)$ decay favors a
negative-parity assignment.
\end{abstract}

\maketitle

The $X(3872)$ meson (denoted in the following as the $X$ meson) has
been observed primarily in its $J/\psi\pi^+\pi^-$ decay
mode~\cite{Choi:2003ue,Acosta:2003zx,Abazov:2004kp,Aubert:2004ns,Aubert:2005zh,Aubert:2008gu}. Evidence
for its decay to the
$J/\psi\gamma$~\cite{Abe:2005ix,Aubert:2006aj,:2008rn} and
$\psi(2S)\gamma$~\cite{:2008rn} final states has established positive
$C$ parity. Analyses by the CDF Collaboration of the $\pi^+\pi^-$ mass
distribution~\cite{Abulencia:2005zc}, and of the decay angular
distribution~\cite{Abulencia:2006ma}, for the $J/\psi\pi^+\pi^-$ decay
mode have narrowed the possible spin-parity ($J^P$) assignment to
$1^+$ or $2^-$. The decay $X\to D^0\bar{D^0}\pi^0$ has also been
observed~\cite{Gokhroo:2006bt} and interpreted as evidence for $X\to
D^{\ast 0}\bar{D^0}$; this has been confirmed by subsequent
analyses~\cite{Aubert:2007rva,:2008su}. There has been much
theoretical interest in the nature of the
\x\/~\cite{Close:2005iz,Close:2007ny,Tornqvist:2004qy,Braaten:2003he,Swanson:2003tb,Voloshin:2004mh,
Voloshin:2007hh,Maiani:2004vq}. Hence, additional experimental
information on new decay modes, especially those sensitive to the
$J^P$ assignment, is germane to the theoretical understanding of this
state.

In a previous \babar\ publication~\cite{Aubert:2007vj}, we have
confirmed the observation of the $Y(3940)$ meson (denoted in the
following as the $Y$ meson) in the decay mode $Y\to J/\psi\omega$
reported by the Belle Collaboration in $B^{0,+}\to J/\psi\omega
K^{0,+}$ decay~\cite{Abe:2004zs}. In the \babar\ analysis of this
decay mode, the $\omega\to\pi^+\pi^-\pi^0$ mass (\mppp\/) region was
defined as $0.7695\leq m_{3\pi}\leq0.7965$ \gevcc\/. With this
requirement and the other selection criteria of
Ref.~\cite{Aubert:2007vj}, we reported no evidence for the decay $X\to
J/\psi\omega$, although Monte Carlo (MC) simulation of $X$-meson decay
to an $S$-wave \jo\ system indicated that this decay could have been
observed. An unpublished Belle analysis of $B^+\to
J/\psi\pi^+\pi^-\pi^0K^+$~\cite{Abe:2005ix}, which required
$\left|m(J/\psi\pi^+\pi^-\pi^0)-3.872\right| <0.0165$ \gevcc\/,
reported evidence for the decay $X\to J/\psi\omega$ on the basis of
$12.4\pm4.1$ events in the mass interval $0.750\leq m_{3\pi}\leq
0.775$ \gevcc\/.

In this study we repeat our analysis of the decay modes $B^{0,+}\to
J/\psi\pi^+\pi^-\pi^0K^{0,+}$~\cite{Aubert:2007vj,conjugate},
extending the selected \mppp\ region to $0.5<$\mppp\/$<0.9$ \gevcc\/
in order to investigate the \mppp\ distribution in a broader region
around the $\omega$ meson.

The data were collected with the \babar\ detector~\cite{Aubert:2001tu}
at the PEP-II asymmetric-energy $e^+e^-$ collider operated at the
$\Upsilon(4S)$ resonance. We use the entire integrated luminosity at
this center-of-mass (c.m.) energy ($\sim 426$ fb$^{-1}$), which yields
a data sample corresponding to about $467\times 10^6$ $B\bar{B}$
pairs. The data were reprocessed with improved charged-particle-track
reconstruction and identification.

The event-selection criteria are identical to those in Table I of
Ref.~\cite{Aubert:2007vj}, except for the initial \mppp\
requirement. The $B$-meson signal region is defined using the
c.m.~energy difference $\DeltaE=E_B^{\ast}-\sqrt{s}/2$, and the
beam-energy substituted mass
$m_{ES}=\sqrt{[(s/2+\vec{p}_i\cdot\vec{p}_B)/E_i]^2-|\vec{p}_B|^{~2}}$~\cite{Aubert:2001tu},
where $(E_i,\vec{p}_i)$ is the initial state four-momentum vector in
the laboratory frame (l.f.), $\sqrt{s}$ is the c.m.~energy,
$E_B^{\ast}$ is the $B$ meson energy in the c.m., and $\vec{p}_B$ is
its l.f.~momentum. Signal $B^+$ ($B^0$) candidates satisfy
$|\DeltaE|<20$ \mev\ (15 \mev\/). In events with multiple $B$
candidates (12$\%$ of events in the region 5.274$<$\mes\/$<$5.284
\gevcc\/), the candidate with the smallest $|\DeltaE|$ is chosen.

\begin{figure}[!htb]
\begin{center}
\includegraphics[width=8.4cm]{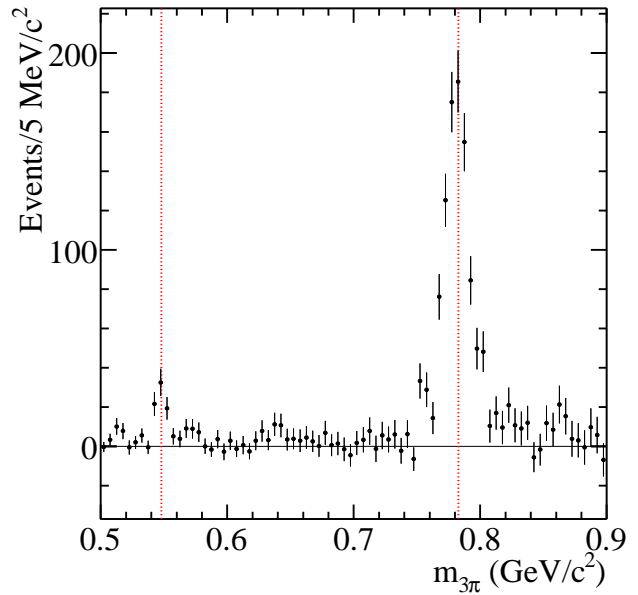}
\caption{The \mppp\ distribution for \bpto\/ candidates obtained as
described in the text; the dashed vertical lines denote the central
mass values of the $\eta$- and $\omega$- meson~\cite{Amsler:2008zzb}.}
\label{fig:1}
\end{center}
\end{figure}

For the $B^+$-candidate sample, the \mppp\ distribution is shown in
Fig.~\ref{fig:1}. The contribution in each mass interval is obtained
by fitting the corresponding \mes\ distribution in the region
$5.2<$\mes\/$<5.3$ \gevcc\/ with a $B^+$ signal Gaussian function and
an ARGUS background function~\cite{Albrecht:1990cs}. The Gaussian mean
value ($\mu$), width ($\sigma$), and the ARGUS parameter (\carg\/),
are fixed to the values obtained when fitting \mes\ for the entire
\jthreepi\ mass region separately for the $B^+$ and $B^0$ samples (for
the $B^+$ sample $\mu=5278.95\pm 0.13$ \mevcc\/, $\sigma=2.83\pm 0.14$
\mevcc\/, and \carg\/$=-37.9\pm 1.8$). A binned Poisson likelihood fit
is performed to the \mes\ distribution in each \mppp\ interval to
obtain the Gaussian and ARGUS normalization parameter values, and
hence to extract the $B$-meson signal.

In Fig.~\ref{fig:1} there is a small, but clear, $\eta$-meson signal,
a large $\omega$-meson signal, and nothing of significance in
between. The $J/\psi\eta$ mass distribution shows no significant
structure, and will not be discussed any further.

In the $\omega$-meson region, the signal extends down to $\sim0.74$
\gevcc\/; there is also a high-mass tail above $\sim0.8$ \gevcc\/, and
possibly some small nonresonant contribution in this region. When we
assign $\omega$-Dalitz-plot weights~\cite{weight} to the events in the
region $0.74-0.80$ \gevcc\/, the sum of weights ($1030\pm 90$) is
consistent with the signal size ($1160\pm 60$), indicating that any
non-$\omega$ background is small, and so we ignore such
contributions. Similar behavior is observed for $B^0$ decay, but with
a selected-event sample which is about six times smaller. In the
following, we define the lower limit of the $\omega$-meson mass region
as 0.74 \gevcc\/, but leave the upper limit at 0.7965 and 0.8055
\gevcc\ for the $B^+$ and $B^0$ samples~\cite{Aubert:2007vj},
respectively, in order to focus on the impact of this one change on
the observed \jo\ mass distribution. The extension of the \mppp\
region towards lower values increases the efficiency slightly.

The \jo\ mass distributions for $B^{0,+}\to J/\psi\omega K^{0,+}$
candidates are obtained by using the same fit procedure used to obtain
the \mppp\ distribution. We then correct the observed signal yields
for selection efficiency. Events corresponding to $B^{0,+}\to
J/\psi\omega K^{0,+}$ decay are created by MC simulation, based on
{\textsc{Geant4}}~\cite{ref:geant}, in order to provide uniform
coverage of the the entire \mjo\ range. The generated events are
subjected to the reconstruction and selection procedures applied to
the data. For $B^+$ ($B^0$) decay it is found that the efficiency
increases (decreases) gradually from $\sim 6\%$ ($\sim 5\%$) close to
\mjo\ threshold to $\sim 7\%$ ($\sim 4\%$) for \mjo\/$\sim 4.8$
\gevcc\/. Comparison of generated and reconstructed \mjo\ values
within each reconstructed \mjo\ mass interval enables the measurement
of the \mjo\ dependence of the mass resolution. From a single-Gaussian
fit to each distribution, the rms deviation is found to degrade
gradually from 6.5 \mevcc\ at \mjo\/$\sim$3.84 \gevcc\/, to 9 \mevcc\
at \mjo\/$\sim$4.8 \gevcc\/.

\begin{figure}[!htp]
\begin{center}
\includegraphics[width=8.4cm]{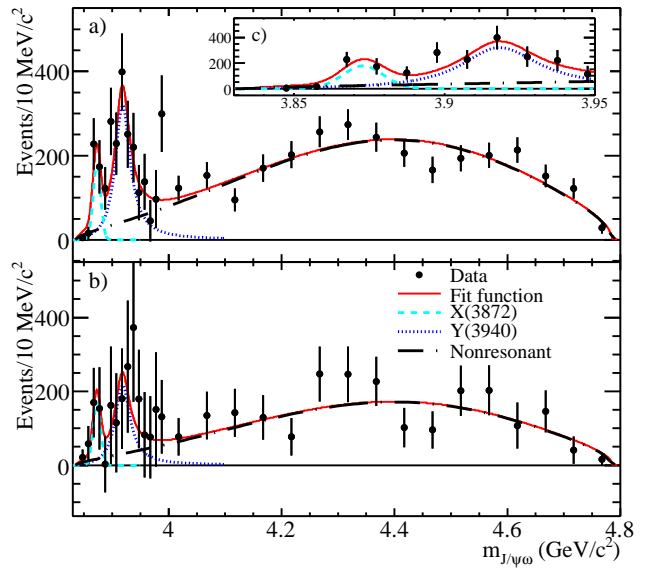}
\caption{The corrected \mjo\ distribution for (a) $B^+$, (b) $B^0$
decays; (c)(inset) shows the low-mass region of (a) in detail. The
curves indicate the results of the fit.}
\label{fig:2}
\end{center}
\end{figure}

The \mjo\ distributions for $B^+\to J/\psi\omega K^+$ and $B^0\to
J/\psi\omega K^0$ decay, after efficiency correction in each mass
interval, are shown in Fig.~\ref{fig:2}(a) and Fig.~\ref{fig:2}(b),
respectively. For the latter, corrections for $K^0_L$ production and
$K^0_S\to\pi^0\pi^0$ decay have been incorporated. The \mjo\ range
from 3.8425 to 3.9925 \gevcc\ is divided into 10 \mevcc\ intervals,
while beyond this 50 \mevcc\ intervals are used. The same choice of
intervals was used in Ref.~\cite{Aubert:2007vj}, where the first two
were inaccessible, and the third was only partly accessible, because
of the value of the lower limit on \mppp\/. Clear enhancements are
observed in the vicinity of the $X$ and $Y$ mesons in the $B^+$
distribution, and similar effects are present in the $B^0$
distribution, with lower statistical significance.

The function used to fit the distributions of Fig.~\ref{fig:2} is a
sum of three components. The \x\ component is a Gaussian resolution
function with fixed rms deviation $\sigma=6.7$ \mevcc obtained from MC
simulation; the intrinsic width of the \x\/ (estimated to be $\lesssim
3$ \mev\/~\cite{Amsler:2008zzb}) is ignored. The $Y$-meson intensity
contribution is represented by a relativistic $S$-wave Breit-Wigner
(BW) function~\cite{Aubert:2007vj}. The nonresonant contribution is
described empirically by a Gaussian function multiplied by \mjo\/. The
$Y$-meson and nonresonant intensity contributions are multiplied by
the phase space factor $p\times q$, where $p$ is the $K$ momentum in
the $B$ rest frame, and $q$ is the $J/\psi$ momentum in the rest frame
of the $J/\psi 3\pi$ system. A simultaneous $\chi^2$ fit to the
distributions of Figs.~\ref{fig:2}(a) and \ref{fig:2}(b) is carried
out, in which only the normalization parameters of the three
contributions are allowed to differ between Fig.~\ref{fig:2}(a) and
Fig.~\ref{fig:2}(b). The fit describes the data well
($\chi^2/\mathrm{NDF} = 54.7/51$, NDF=number of degrees of freedom),
as shown by the solid curves in Fig.~\ref{fig:2}. The dashed and
dotted curves show the $X$- and $Y$-meson contributions, respectively,
while the dot-dashed curves represent the nonresonant distribution.

For the \x\/, the fitted mass is $3873.0_{-1.6}^{+1.8}\stat \pm
1.3\syst $ \mevcc\/, while the mass and width values for the \y\ are
$3919.1_{-3.4}^{+3.8}\stat \pm 2.0\syst $ \mevcc\ and
$31_{-8}^{+10}\stat \pm 5\syst $ \mev\/, respectively.  These results
are consistent with earlier \babar\
measurements~\cite{Aubert:2008gu,Aubert:2007vj}.

From the fits of Fig.~\ref{fig:2}, we obtain product branching
fraction measurements for $B^{0,+}\to X K^{0,+}$, \xto\/. The
resulting $B^+$ and $B^0$ product branching fraction values are
$[0.6\pm 0.2\stat \pm 0.1\syst ]\times 10^{-5}$, and $[0.6\pm 0.3\stat
\pm 0.1\syst ]\times 10^{-5}$, respectively.

Similarly, we obtain updated values for ${\cal{B}}(B^+\to YK^+)\times
{\cal{B}}(Y\to J/\psi\omega) = [3.0_{-0.6}^{+0.7}\stat
_{-0.3}^{+0.5}\syst ]\times 10^{-5}$, ${\cal{B}}(B^0\to YK^0)\times
{\cal{B}}(Y\to J/\psi\omega) = [2.1\pm 0.9\stat \pm 0.3\syst ]\times
10^{-5}$, and for the total ({\it{i.e.}} the sum of the $X$- meson,
$Y$-meson, and nonresonant, contributions) ${\cal{B}}(B^+\to
J/\psi\omega K^+) = [3.2\pm 0.1\stat _{-0.3}^{+0.6}\syst ]\times
10^{-4}$ and ${\cal{B}}(B^0\to J/\psi\omega K^0)=[2.3\pm 0.3\stat \pm
0.3\syst ]\times 10^{-4}$. These values are consistent with those of
Ref.~\cite{Aubert:2007vj}, and supersede them.

We define $R_X$, $R_Y$, and $R_{NR}$ as the ratios of the $B^0$ to
$B^+$ branching fractions to the final states $XK$, $YK$, and
nonresonant $J/\psi\omega K$, and extract these ratios from a
simultaneous fit to the data, with the fit function adjusted to
explicitly contain these parameters. This yields
$R_X=1.0_{-0.6}^{+0.8}\stat _{-0.2}^{+0.1}\syst $,
$R_Y=0.7_{-0.3}^{+0.4}\stat \pm 0.1\syst $, and $R_{NR}=0.7\pm
0.1\stat \pm 0.1\syst $. The values of $R_Y$ and $R_{NR}$ are
consistent with those in Ref.~\cite{Aubert:2007vj}. The statistical
uncertainty on $R_{NR}$ has been reduced significantly with respect to
Ref.~\cite{Aubert:2007vj} as a result of the increased luminosity,
improvements in event reconstruction efficiency, and the use of much
larger MC samples in the measurement of the selection efficiency as a
function of \mjo\/, especially for \mjo\/$>4$ \gevcc\/.

In Ref.~\cite{Aubert:2008gu}, it was found that ${\cal{B}}(B^+\to
XK^+)\times{\cal{B}}(X\to J/\psi\pi^+\pi^-)=[8.5\pm 1.5\stat \pm
0.7\syst ]\times 10^{-6}$ and ${\cal{B}}(B^0\to
XK^0)\times{\cal{B}}(X\to J/\psi\pi^+\pi^-)=[3.5\pm 1.9\stat \pm
0.4\syst ]\times 10^{-6}$. We combine these results with those from
the present analysis to obtain the ratio of the branching fractions
${\cal{B}}(X\to J/\psi\omega)/{\cal{B}}(X\to J/\psi\pi^+\pi^-)$. For
$B^+$ ($B^0$) events, this ratio is $0.7\pm 0.3$ ($1.7\pm 1.3$), where
the statistical uncertainties, and those systematic uncertainties
which do not cancel in the ratio, have been added in quadrature; the
weighted average is $0.8\pm 0.3$. This is consistent with that
reported in Ref.~\cite{Abe:2005ix} ($1.0\pm0.4\stat \pm 0.3\syst $).

In obtaining the quoted systematic errors, systematic uncertainties
due to tracking ($2\%$), particle identification (4.4$\%$ and 5.2$\%$
for $B^0$ and $B^+$ events), $\pi^0$ reconstruction efficiency
(3.6$\%$), $K^0_S$ reconstruction efficiency (2$\%$) for the $B^0$
events, and $B\bar{B}$ event counting (1.1$\%$), have been taken into
account. The uncertainties on the branching fraction values for
$J/\psi\to\ell^+\ell^-$ and $\omega\to 3\pi$~\cite{Amsler:2008zzb}
have been treated as sources of systematic uncertainty. When fitting
the \mes\ distributions in each \mjo\ or \mppp\ mass interval, the
parameters $\mu$, $\sigma$, and \carg\ were fixed to the values
obtained from the fit to the corresponding total \mes\
distribution. Associated systematic uncertainties were estimated by
increasing and decreasing the central value of each parameter by one
standard deviation, repeating the analysis, and taking the change in
each fitted quantity as an estimate of systematic
uncertainty. Similarly, the systematic uncertainty associated with the
efficiency-correction procedure was estimated by varying its \mjo\
dependence within a $\pm 1\sigma$ envelope, repeating the fits to the
data of Fig.~\ref{fig:2}, and taking the corresponding changes in fit
parameter values as estimates of systematic uncertainty. Additional
systematic uncertainties on the mass and width of the \y\ were
estimated as described in Ref.~\cite{Aubert:2007vj}. The main
contributions described there result from a comparison of the MC input
values to those obtained after event reconstruction, and from the
difference in fitted values when a $P$-wave BW was used instead of an
$S$-wave BW to describe the $Y$-meson lineshape.

Since the $X$-meson signal occurs at a low statistical level and at
very low values of \mjo\/, there is concern that the measured
signal-event yield might be biased because of the low mass tails of
the $Y$-meson and nonresonant contributions. A detailed MC study using
samples of $X$-meson events ranging in size from 10-500 events showed
no evidence of bias, and the spread in extracted signal yield was
consistent with the corresponding statistical uncertainty obtained
from the fit to the data.

\begin{figure}[!htb]
\begin{center}
\includegraphics[width=8.4cm]{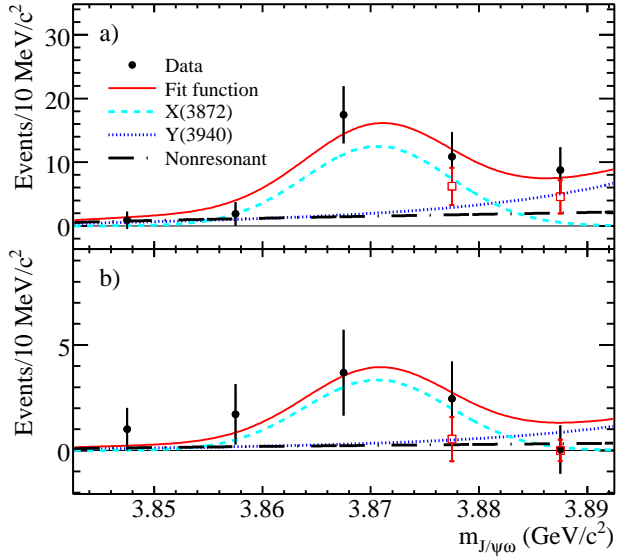}
\caption{The uncorrected \mjo\ distributions for events with
$3.8425<$\mjo\/$<3.8925$ \gevcc\ for (a) $B^+$ and (b) $B^0$ decays;
the open squares correspond to (a) \mppp\/$>0.7695$ (b)
\mppp\/$>0.7605$ \gevcc\/~\cite{Aubert:2007vj}. The curves indicate
the results of the fit.}
\label{fig:3}
\end{center}
\end{figure}

\begin{figure}[!htb]
\begin{center}
\includegraphics[width=8.4cm]{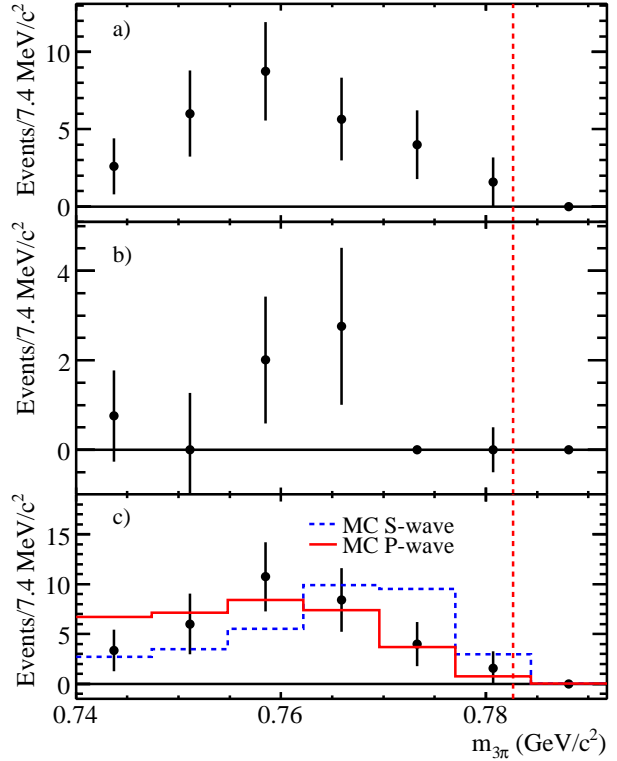}
\caption{The \mppp\ distribution for events with
$3.8625<$\mjo\/$<3.8825$ \gevcc\ for (a) $B^+$, (b) $B^0$, and (c) the
combined distribution. The vertical dashed line indicates the nominal
$\omega$-meson mass~\protect\cite{Amsler:2008zzb}. In (c), the solid
(dashed) histogram represents reconstructed MC $P$-wave ($S$-wave)
events normalized to the number of data events.}
\label{fig:4}
\end{center}
\end{figure}

We now consider the relationship between the $X$-meson signal and the
choice of lower mass limit for the $\omega$-meson region. In
Fig.~\ref{fig:3} we show the data corresponding to the first five mass
intervals of Fig.~\ref{fig:2} (3.8425$<m_{J/\psi\omega}<$3.8925
\gevcc\/) before applying the efficiency and $K^0$ branching fraction
corrections. The points shown by open squares indicate the effect of
choosing the \mppp\ lower limit to be 0.7695 \gevcc\ rather than 0.740
\gevcc\/.  The three lowest intervals then yield no signal, and the
other two contain only 11 (0.5) events in Fig.~\ref{fig:3}(a)
(Fig.~\ref{fig:3}(b)). This is to be compared with
$42.4_{-7.2}^{+7.8}$ ($8.5_{-3.0}^{+3.7}$) events obtained when the
\mppp\ lower limit is 0.74 \gevcc\/. Since the number of events in
Fig.~\ref{fig:3} is much smaller than the total number of
$\omega$-meson events ($1160\pm 60$ for $B^+$ and $206\pm 26$ for
$B^0$ decay), and since the \mppp\ distribution (Fig.~\ref{fig:4}(c))
differs greatly from the $\omega$-meson lineshape, these might be
nonresonant $3\pi$ events. To check the $\omega$-meson interpretation,
we sum the $\omega$-Dalitz-plot weights~\cite{weight} for the events
contributing to Fig.~\ref{fig:3}(a) (solid points) in the \mes\ signal
region and obtain $41\pm 13$, in good agreement with the number from
the \mes\ fits. This justifies the $\omega$-meson interpretation. In
contrast, we note that for the $152\pm 20$ $\eta$-meson events in
Fig.~\ref{fig:1} the sum of the weights~\cite{weight} is $-1\pm 42$,
as expected for a uniform Dalitz-plot distribution.

To determine the significance of the $X\to J/\psi\omega$ signal, we
extract the signal yields from a fit to the data, prior to the
corrections for efficiency and $K^0$ branching fractions, as shown in
Fig.~\ref{fig:3}. The fitted values of the masses and widths are in
agreement with those obtained from the fit to the corrected data. An
$X$-meson signal of $21.1\pm 7.0$ events is obtained for $B^+$ decay,
and $5.6\pm 3.0$ events for $B^0$ decay, so that the combined signal
is $26.7\pm 7.6$ events. For the combined distribution, the mass
region $3.8625-3.8825$ \gevcc\ contains $34.0\pm 6.6$ events, and the
fitted curves indicate that only $ 8.9\pm 1.0$ events are due to the
tails of the $Y$-meson and nonresonant distributions. We convolve a
Gaussian ensemble of background Poisson distributions with a Gaussian
distribution of observed events, and obtain probability 3.6$\times
10^{-5}$ that the $34.0\pm 6.6$ events can result from upward
background fluctuation. This corresponds to a significance of
$4.0\sigma$ for a normal distribution. On this basis we report
evidence for the decay mode \xto\/.

For the $3.8625-3.8825$ \gevcc\ region of Fig.~\ref{fig:3}, we plot
the \mppp\ distributions in Fig.~\ref{fig:4}. Each data point results
from a fit to the corresponding \mes\ distribution; for the points
with no error bars, the \mes\ distribution is empty. For the combined
distribution, Fig.~\ref{fig:4}(c), $\sim 84\%$ of the events have
\mppp\/$<0.7695$ \gevcc\/, the mass limit used in
Ref.~\cite{Aubert:2007vj}. The dashed histogram in Fig.~\ref{fig:4}(c)
results from normalizing the reconstructed $X$-meson MC events to the
observed 34 events. Since the \jo\ system was generated with zero
orbital angular momentum, this corresponds to positive $X$-meson
parity. One unit of orbital angular momentum creates a centrifugal
barrier factor $q^2/(1+R^2q^2)$ in the description of the \jo\ final
state, where $R=3$ GeV$^{-1}$ is the $P$-wave Blatt-Weisskopf barrier
factor radius~\cite{Blatt} (values in the range $0<R<5$ GeV$^{-1}$
yield no significant difference). This factor suppresses the \threepi\
mass spectrum near the upper kinematic limit, as shown by the solid
histogram of Fig.~\ref{fig:4}(c) (also normalized to 34 events). For
the dashed histogram $\chi^2/\mathrm{NDF}=10.17/5$ and the
$\chi^2$-distribution probability is $P(\chi^2,\mathrm{NDF})=7.1\%$,
while for the solid histogram $\chi^2/\mathrm{NDF}=3.53/5$ and
$P(\chi^2,\mathrm{NDF})=61.9\%$. It follows that the observed
distribution favors the $P$-wave description both quantitatively and
qualitatively. If both histograms are normalized to the region
\mppp\/$<0.7695$ \gevcc\ (which was excluded in
Ref.~\cite{Aubert:2007vj}), we expect for \mppp\/$>0.7695$ \gevcc\/,
and hence for the \mjo\ interval $3.8725-3.8825$ \gevcc\/, $\sim 4.3$
events for the $P$-wave description, and $\sim 16.6$ events for the
$S$-wave description. However, in Fig.~\ref{fig:3} we observe $\sim 6$
events. In Ref.~\cite{Dunwoodie:2007be}, it was pointed out that for
$X(3872)\to D^{\ast 0}\bar{D^0}$, the introduction of one unit of
orbital angular momentum in the final state could explain the shift in
measured $X$-meson mass~\cite{Gokhroo:2006bt,Aubert:2007rva}. This
observation and the present analysis, together with the spin-parity
($J^P$) analysis of Ref.~\cite{Abulencia:2006ma}, favor $J^P=2^-$ for
the $X(3872)$ meson. For $I=0$ and $J^{PC}=2^{-+}$, the $X$-meson mass
falls within the broad range of estimates for the $\eta_{c2}(1D)$
charmonium state~\cite{Godfrey:1985xj,Barnes:2003vb}. We conclude that
this interpretation is favored by the data.

In summary, we have used the entire \babar\ data sample collected at
the $\Upsilon(4S)$ resonance to obtain evidence for \xto\ in
$B^{0,+}\to J/\psi\omega K^{0,+}$ with product branching fraction
values $[0.6\pm 0.2\stat \pm 0.1\syst ]\times 10^{-5}$ and $[0.6\pm
0.3\stat \pm 0.1\syst ]\times 10^{-5}$ for $B^+$ and $B^0$,
respectively.  A comparison of the observed \mppp\ mass distribution
from \xto\ decay to those from MC simulations leads us to conclude
that the inclusion of one unit of orbital angular momentum in the
$J/\psi\omega$ system significantly improves the description of the
data. This in turn implies negative parity for the \x\/, and hence
$J^P=2^-$ is preferred ~\cite{Abulencia:2006ma}. In addition, we have
updated the mass and width of the \y\ ($3919.1_{-3.5}^{+3.8}\stat \pm
2.0\syst $ \mevcc\ and $31_{-8}^{+10}\stat \pm 5\syst $ \mev\/), the
product branching fraction values for $B^{0,+}\to YK^{0,+}$, \yto\/,
and our measurements of the total branching fractions for $B^{0,+}\to
J/\psi\omega K^{0,+}$.

We are grateful for the excellent luminosity and machine conditions
provided by our \pep2\ colleagues, 
and for the substantial dedicated effort from
the computing organizations that support \babar.
The collaborating institutions wish to thank 
SLAC for its support and kind hospitality. 
This work is supported by
DOE
and NSF (USA),
NSERC (Canada),
CEA and
CNRS-IN2P3
(France),
BMBF and DFG
(Germany),
INFN (Italy),
FOM (The Netherlands),
NFR (Norway),
MES (Russia),
MICIIN (Spain),
STFC (United Kingdom). 
Individuals have received support from the
Marie Curie EIF (European Union),
the A.~P.~Sloan Foundation (USA)
and the Binational Science Foundation (USA-Israel).


\begin{thebibliography}{10}
\bibitem{conjugate} The use of charge conjugate reactions is implied
throughout this Paper.
\bibitem{weight} Each event is given weight
$\frac{5}{2}(1-3\cos^2\theta_h)$, where $\theta_h$ is the angle
between the $\pi^+$ and $\pi^0$ directions in the $\pi^+\pi^-$ rest
frame.
\end{thebibliography}
\end{document}